\documentclass{article}
\usepackage[preprint, nonatbib]{neurips_2023}
\usepackage{graphicx} 
\usepackage{subcaption}
\usepackage{caption}
\usepackage{float}
\usepackage{hhline}
\usepackage{multirow}
\usepackage{appendix}
\usepackage{fancyvrb}
\graphicspath{ {./images/} }

\usepackage[utf8]{inputenc} 
\usepackage[T1]{fontenc}    
\usepackage{hyperref}       
\usepackage{url}            
\usepackage{booktabs}       
\usepackage{amsfonts}       
\usepackage{nicefrac}       
\usepackage{microtype}      
\usepackage{xcolor}         

\title{Strategic Behavior of Large Language Models: Game Structure vs. Contextual Framing}

\author{%
  Nunzio Lorè\\
  Network Science Institute \\
Multi-Agent Intelligent Complex Systems (MAGICS) Lab\\
  Northeastern University, Boston, Massachusetts, USA \\
  \texttt{lora.n@northeastern.edu} \\
    \And
  Babak Heydari\thanks{Corresponding author} \\
  College of Engineering and Network Science Institute \\
  Multi-Agent Intelligent Complex Systems (MAGICS) Lab\\
  Northeastern University, Boston, Massachusetts, USA \\
  \texttt{b.heydari@northeastern.edu} \\
}

\begin{document}

\maketitle

\begin{abstract}
  This paper investigates the strategic decision-making capabilities of three Large Language Models (LLMs): GPT-3.5, GPT-4, and LLaMa-2, within the framework of game theory. Utilizing four canonical two-player games—Prisoner's Dilemma, Stag Hunt, Snowdrift, and Prisoner's Delight—we explore how these models navigate social dilemmas, situations where players can either cooperate for a collective benefit or defect for individual gain. Crucially, we extend our analysis to examine the role of contextual framing, such as diplomatic relations or casual friendships, in shaping the models' decisions. Our findings reveal a complex landscape: while GPT-3.5 is highly sensitive to contextual framing, it shows limited ability to engage in abstract strategic reasoning. Both GPT-4 and LLaMa-2 adjust their strategies based on game structure and context, but LLaMa-2 exhibits a more nuanced understanding of the games' underlying mechanics. These results highlight the current limitations and varied proficiencies of LLMs in strategic decision-making, cautioning against their unqualified use in tasks requiring complex strategic reasoning.
\end{abstract}

\section{Introduction}

Large Language Models (LLMs) such as GPT from OpenAI and LLaMa-2 from Meta have garnered significant attention for their ability to perform a range of human-like tasks that extend far beyond simple conversation. Some argue that these models may serve as an intermediate step toward Artificial General Intelligence (AGI) \cite{bubeck2023sparks}. Recent advancements have shown GPT-4 passing the bar exam \cite{katz2023gpt} and GPT-3 solving complex mathematical problems \cite{zong2023solving}. Despite these achievements, these models exhibit limitations, notably in tasks like network structure recognition \cite{guo2023gpt4graph}.

Social and behavioral science research on Large Language Models (LLMs), including GPT and LLaMa-2, is divided into two principal streams: one that explores human-like cognitive capabilities such as reasoning and theory of mind \cite{arkoudas2023gpt, frith2005theory, singh2023mind, hagendorff2023human, liu2023evaluating}, and another that evaluates performance in comparison to human skills across a variety of tasks \cite{ali2023performance, lin2023comparison, de2023can}. In the field of economics, the emphasis is predominantly on performance evaluation, exploring applications like market research and sentiment analysis \cite{brand2023using, mahdavi2023econsentgpt, chen2023emergence}. This dual focus coalesces in social science research, where LLMs have gained attention for their potential to simulate human behavior in experimental settings \cite{aher2022using, horton2023large, hagendorff2023machine, argyle2023out}. Notably, within the intricate framework of social dilemmas and game theory, LLMs are being tested for both their cognitive reasoning skills and performance outcomes \cite{phelps2023investigating, guo2023gpt, brookins2023playing, akata2023playing}.

Existing studies indicate that LLMs can mimic human behavior to some extent \cite{brookins2023playing, guo2023gpt}, yet their aptitude for strategic decision-making in game-theoretic contexts is still an area for exploration. Beyond the structural elements of a game, the contextual framing can significantly affect decision-making processes. Prior research on human behavior has underlined the powerful role of context in shaping strategic choices; for example, the framing of a game as a Wall Street venture versus a community endeavor led to divergent decisions \cite{liberman2004name}. As a result, our study aims to go beyond assessing the fundamental strategic capabilities of LLMs, also considering the influence of game structure and contextual framing on their decision-making.

To disentangle the complexities of strategic decision-making in LLMs, we conduct a series of game-theoretic simulations on three distinct models: GPT-3.5, GPT-4, and  LLaMa-2. We focus on social dilemmas, games in which players may either cooperate for collective benefit or defect for individual gain. Starting from the well-known Prisoner's Dilemma, we expand our study to include other two-player games such as the Stag Hunt, Snowdrift, and Prisoner's Delight (aka Harmony Game). Besides examining these games, we introduce five different contexts—ranging from business and diplomatic discussions to casual interactions between friends—to evaluate how contextual framing influences strategic choices. Our primary research question is to determine the relative significance of game structure versus contextual framing in shaping the behavior of these models. 

Our findings unveil the subtle intricacies in how each of the examined Large Language Models responds to strategic scenarios. GPT-3.5 appears particularly sensitive to contextual framing but demonstrates limited proficiency in grasping abstract strategic considerations, such as reasoning based on a \textit{best response} strategy. In contrast, both GPT-4 and LLaMa-2 exhibit a more balanced approach, adjusting their strategies based on both the intrinsic game structure and the contextual framing. Notably, the impact of context is more pronounced in specific domains, such as interactions framed as games among friends, where the game structure itself takes a backseat.

When it comes to comparing GPT-4 and LLaMa-2, our findings reveal that GPT-4, on average, places greater weight on the game structure than on context, relative to LLaMa-2. However, prioritizing game structure over context does not translate to a nuanced differentiation between distinct game types. In fact, GPT-4 seems to employ a binary threshold approach, categorizing games into 'high' and 'low' social dilemma buckets, rather than discerning the unique features of each game. Contrary to this, LLaMa-2 exhibits a more finely-grained understanding of the various game structures, even though it places greater emphasis on contextual factors compared to GPT-4. This suggests that LLaMa-2 is better equipped to navigate the subtleties of different strategic scenarios while also incorporating context into its decision-making, whereas GPT-4 adopts a more generalized, structure-centric strategy.

In addition to analyzing the decision-making patterns of these large language models, we examined anecdotal evidence to further decipher the mechanisms behind their distinct behaviors. GPT-3.5 appears to have a rudimentary understanding of strategic scenarios, frequently failing to identify best responses and committing a variety of basic mathematical errors. GPT-4, on the other hand, demonstrates a higher level of sophistication in its arguments. It often begins its reasoning by modeling the game structure and conditioning its responses based on anticipated actions of other players. However, GPT-4 also tends to mischaracterize game structures, often reducing them to variations of the Prisoner's Dilemma, even when the structural nuances suggest otherwise. Interestingly, it adopts a different line of reasoning in games framed between friends, emphasizing the importance of longer-term relationships over immediate payoff maximization—despite explicit game descriptions to the contrary. LLaMa-2 approaches these strategic scenarios differently, initially abstracting the problem to a higher level using explicit game-theoretic language. It then layers contextual elements on top of this game-theoretic foundation, offering a well-rounded analysis that encompasses both game structure and situational factors.



\section{Methods}

\begin{figure}[H]
  \includegraphics[scale=0.55]{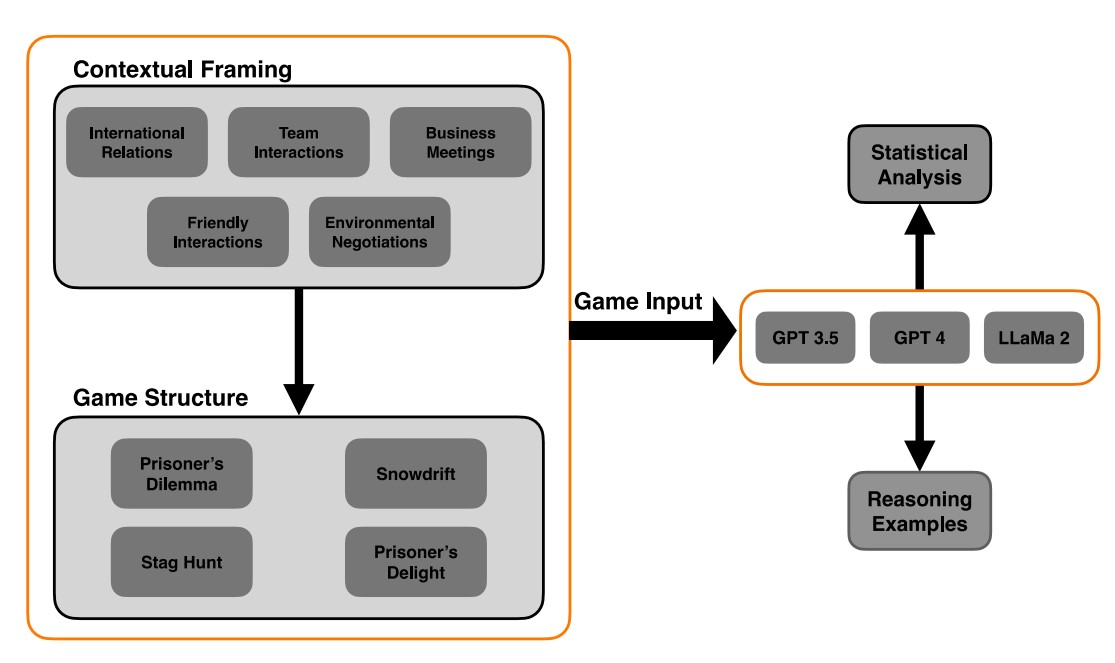}
  \caption{A schematic explanation of our data collecting process. A combination of a contextual prompt and a game prompt is fed into one of the LLM we examine in this paper, namely GPT-3.5, GPT-4, and LLaMa-2. Each combination creates a unique scenario, and for each scenario we collect 300 initializations. The data for all scenarios played by each algorithm is then aggregated and used for our statistical analysis, while the motivations provided are scrutinized in our Reasoning Exploration section.}
  \label{process}
\end{figure}

Figure \ref{process} shows the schematic workflow of this research and the process through which we generate our results. To each game we combine a context, a term we use to indicate the social environment in which the interaction described by the model takes place. We run 300 initializations per LLM for each of the 20 possible unique combinations of context and game, before aggregating the results in order to conduct our statistical analysis. \\

We run our experiments using OpenAI's \verb|gpt-3.5-turbo-16k| and \verb|gpt-4| models, interfacing with them through Python's \verb|openai| package. For LLaMa-2, we utilize Northeastern University's High Performance Cluster (HPC) as the model lacks a dedicated API or user interface. We access LLaMa-2 via the HuggingFace pipeline. To standardize our simulations, we restrict the response token count to $50$ for the OpenAI models and $8$ for LLaMa-2, setting the temperature parameter at 0.8. We opt for this temperature setting for several reasons: first, it mirrors the default settings in user-based applications like chatGPT, providing a realistic baseline; second, it allows for the exploration of multiple plausible actions in games with mixed Nash equilibria; and third, lower temperature settings risk obscuring the inherently probabilistic nature of these algorithms and may produce unengaging results. We note that high temperatures are commonly used in related working papers \cite{kojima2022large, wei2022chain}.

Our experimental design includes two distinct prompts for each LLM. The initial prompt sets the context, outlining the environment and directing the algorithm to assume a specific role. Its aim is to create a realistic setting for the game to take place. The second prompt establishes the "rules," or more accurately, the payoff structure of the game. While contextual prompts are disseminated via the system role, the payoff prompts are communicated through the user role. In both scenarios, we adhere to best practices such as advising the model to deliberate thoughtfully and utilizing longer prompts for clarity \cite{kojima2022large, wei2022chain}. The contextual prompts are crafted to be universally applicable to the range of games examined, sacrificing some degree of specificity for broader relevance. Detailed text for each prompt is available in Appendix A. Summarizing, we present the following scenarios:

\begin{itemize}
    \item A summit between two heads of state from two different countries ("IR"),
    \item A meeting between two CEOS from two different firms ("biz"),
    \item A conference between two industry leaders belonging to two different companies making a joint commitment on environmental regulations ("environment"),
    \item A talk between two employees who belong to the same team but are competing for a promotion ("team"),
    \item A chat between two friends trying to reach a compromise ("friendsharing").
\end{itemize}

The games we use for our analysis are borrowed from the literature on social dilemmas in game theory. In particular, they all have the following form:

\begin{table}[h]
    \centering
    \begin{tabular}{|c|c|c|}
    \hline
      & C & D \\
    \hline
    C & (R,R) & (S, T) \\
    \hline
    D & (T, S) & (P,P) \\
    \hline
    \end{tabular}
    \label{tab:my_label}
\end{table}

In this paper, we define "social dilemmas" any strategic interaction models that feature two types of actions: a socially optimal action that benefits both players if chosen mutually, and an individually optimal action that advantages one player at the expense of the other. We refer to the socially optimal action as "cooperation," abbreviated as "C," and the individually optimal action as "defection," also abbreviated as "D." For clarity, each pair of actions taken by players corresponds to a payoff vector, which we express in terms of utils or points, following standard game theory conventions. The first entry in the vector represents the row player's payoff, while the second entry is reserved for the column player. In this framework, "R" signifies the reward for mutual cooperation, "T" represents temptation to defect when the other player cooperates, "S" indicates the sucker's payoff for cooperating against a defector, and "P" stands for the punishment both players receive when both choose to defect, typically leading to a suboptimal outcome for both. Different relationships between these values give rise to different games:

\begin{itemize}
    \item When T $>$ R $>$ P $>$ S, the game is the Prisoner's Dilemma;
    \item When T $>$ R $>$ S $>$ P, the game is Snowdrift, also known as Chicken;
    \item When R $>$ T $>$ P $>$ S, the game is Stag Hunt;
    \item When R $>$ T $>$ S $>$ P, the game is the Prisoner's Delight, also known as Harmony. 
\end{itemize}

This structure is in the spirit of \cite{gianetto2013catalysts} and \cite{gianetto2015network}, in which the same four game theoretic models are used to capture different types and degrees of social dilemma. We point out that Prisoner's Delight is not exactly a dilemma, but rather an anti-dilemma, as choosing to cooperate is both socially and individually optimal. On the opposite end of the spectrum lies the Prisoner's Dilemma, in which defecting is always optimal and thus leads to a situation in which both players are worse off, at least according to standard predictions in Game Theory.

Here we introduce a piece of important terminology: in the Prisoner's Dilemma and in the Prisoner's Delight, only one action is \textit{justifiable}. This means that one action strictly dominates another, and therefore a rational player would only ever play the strictly dominant action. The Stag Hunt and Snowdrift lie somewhere in between, with both cooperation and defection being justifiable. More specifically, in the Stag Hunt, the Nash Equilibrium in pure actions is reached if both players coordinate on the same action (with the cooperative equilibrium being payoff dominant), whereas in Snowdrift said equilibrium is reached if both players coordinate on opposite actions. As neither action strictly dominates the other, a rational player is justified in playing either or both, and in fact for these games an equilibrium exists in mixed strategies as well.  \\

For each game and for each context, we run 300 initializations and record the action taken by the LLM agent, and keep track of the rate of cooperation by the LLM agents for our follow up analysis. For each experiment, we keep the prompts constant across LLMs. 

\section{Results}

\begin{figure}[h]
  \subcaptionbox{Results grouped game, GPT-3.5}{\includegraphics[width = 2.7in]{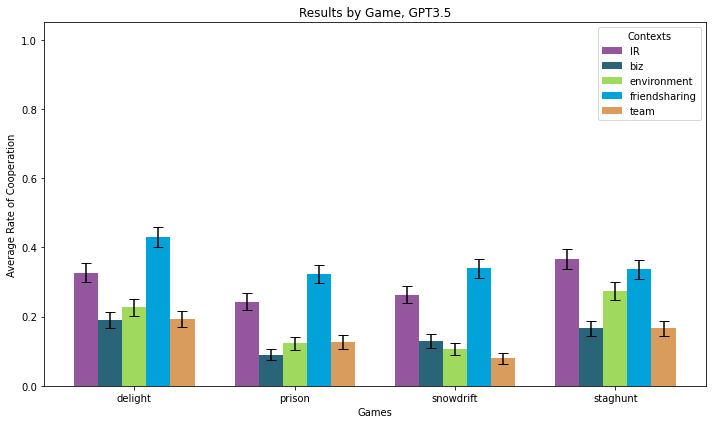}}\quad
  \subcaptionbox{Results grouped by context, GPT-3.5}{\includegraphics[width = 2.7in]{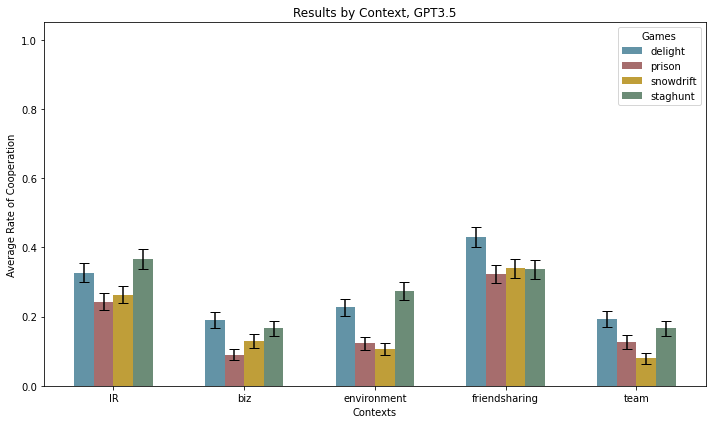}}
  \subcaptionbox{Results grouped by game, GPT-4}{\includegraphics[width = 2.7in]{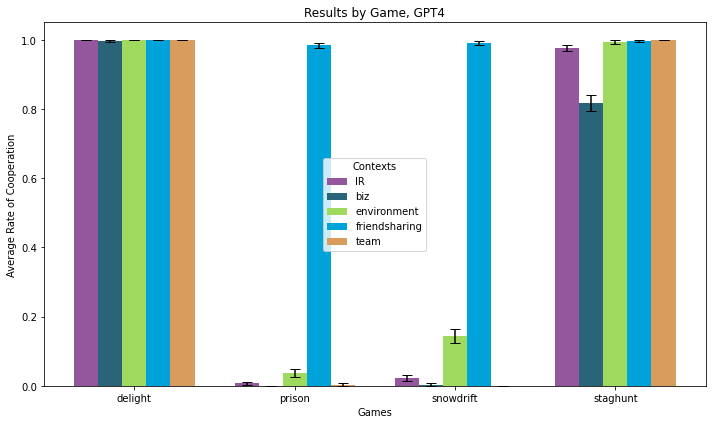}}\quad
  \subcaptionbox{Results grouped by context, GPT-4}{\includegraphics[width = 2.7in]{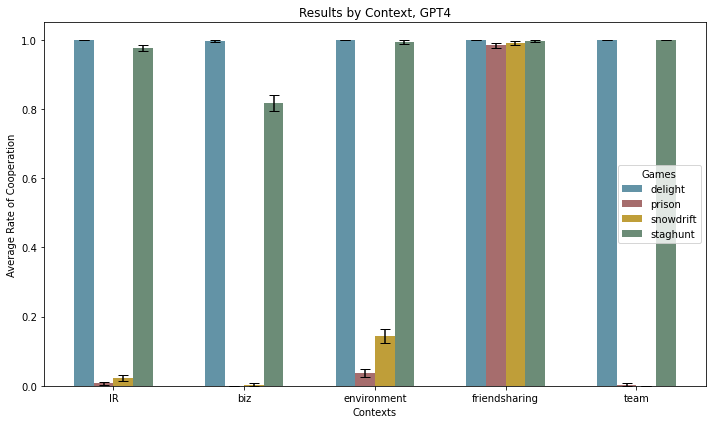}}
  \subcaptionbox{Results grouped by game, LLaMa-2}{\includegraphics[width = 2.7in]{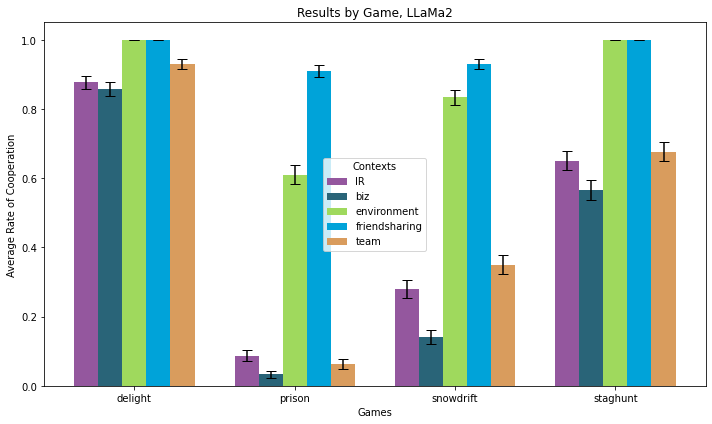}}\quad
  \subcaptionbox{Results grouped by context, LLaMa-2}{\includegraphics[width = 2.7in]{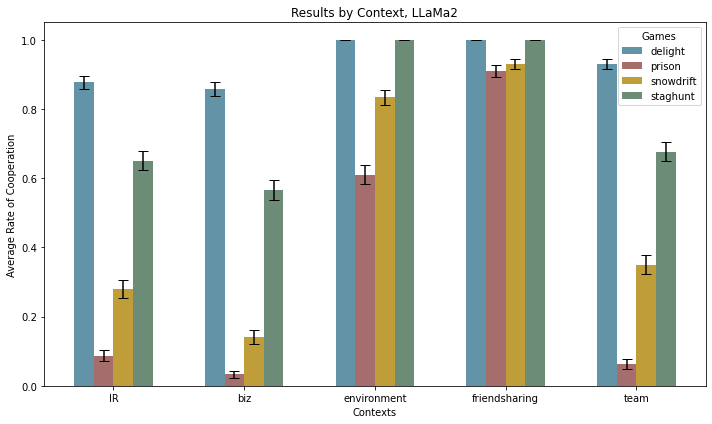}}
  \caption{Summary of our findings, displayed using bar charts and outcomes grouped either by game or by context. On the $y$ axis we display the average propensity to cooperate in a given game and under a given context, with standard error bars. Figures (a) and (b) refer to our experiments using GPT-3.5, and anticipate one of our key findings: context matters more than game in determining the choice of action for this algorithm. Figures (c) and (d) instead show how the opposite is true for GPT-4: almost all contexts are more or less playing the same strategy, that of cooperating in two of the four games and defecting in the remaining two. Finally, Figures (e) and (f) present our results for LLaMa-2, whose choice of action clearly depends both on context and on the structure of the game.}
  \label{figalpha}
\end{figure}

Figure \ref{figalpha} displays an overview of our results for all three LLMs. To better clarify the role of game structure vs. framing context, results are aggregated at different levels: we group the observations at the game level on the left at the context level on the right, and each row represents a different LLM. A few things appear immediately clear when visually inspecting the figure. First, GPT-3.5 tends not to cooperate regardless of game or context. Second, GPT-4's choice of actions is almost perfectly bimodal, with either full cooperation or full defection. Finally, LLaMa-2's behavior approximates that of GPT-4 to a certain extent, but with a wider degree of variation between response both across games and across contexts. A more detailed view of strategical choice for each game, context and LLM is presented in Appendix B. \\ 

To further corroborate and substantiate our findings, we turn to dominance analysis using STAT. In practice, dominance analysis is used to study how the prediction error changes when a given independent variable is omitted from a statistical model. This procedure generates $2^{x}-1$ nested models, with $x$ being the number of regressors. The larger the increase on average over the nested models in error, the greater the importance of the predictor. \cite{doi:10.1177/1536867X211025837}. We run a logit regression for each LLM encoding each game and each context as a dummy variable, and then we use dominance analysis to identify which dummies have the largest impact on the dependant variable. The output is presented in Table \ref{domtable}. We notice that "friendsharing" consistently ranks in the top spots across all algorithms, and indeed by analyzing Figure \ref{figalpha} it appears immediately clear that this context is consistently associated with higher rates of cooperation regardless of game or LLM. For GPT-3.5, contexts represent the five most important variables, with games with a sole rationalizable action occupying positions 6 and 7. This suggests that GPT-3.5 might have a tendency to put weight on context first and on game structure last, with a slight bias for "simpler" games. For GPT-4, on the other hand, the ranking is almost perfectly inverted with games being the regressors with the highest dominance score. Prisoner's Delight and Dilemma once again rank the highest among games for influence, while "friendsharing" is dethroned and relegated to the second position. The ranking for LLaMa-2 paints a more nuanced picture, with contexts and games alternating throughout the ranking, but with "friendsharing" still firmly establishing itself as the most influential variable. \\

\begin{table}[h]
\begin{center}
\begin{tabular}{{c}{c}{c}{c} }
\hline\hline
        Independent Variable & GPT3.5 & GPT-4 & LLaMa-2 \\
\hline
Prisoner's Dilemma      &  0.00266  &  0.160  & 0.0810       \\
Stag Hunt    &   0.00201 &  0.143  & 0.0281      \\
Snowdrift   &   0.00240  & 0.142   &  0.0278    \\
Prisoner's Delight     &   0.00298  & 0.186  & 0.0883        \\
IR          &   0.00646  & 0.0193   & 0.0283      \\
biz         &   0.00762 &  0.0320   & 0.0467       \\
friendsharing&   0.0156 &  0.161 &  0.0909      \\
environment &    0.00316   & 0.0146  & 0.0438       \\
team        &    0.00803  & 0.0194  &  0.0225    \\
\hline
\(N\)       &   6000  &  6000  & 6000       \\
\hline\hline
\end{tabular}
\caption{Results of the dominance analysis for each LLM.} \label{domtable}
\end{center}
\end{table}

\begin{figure}[!h]
 \includegraphics[scale = 0.5]{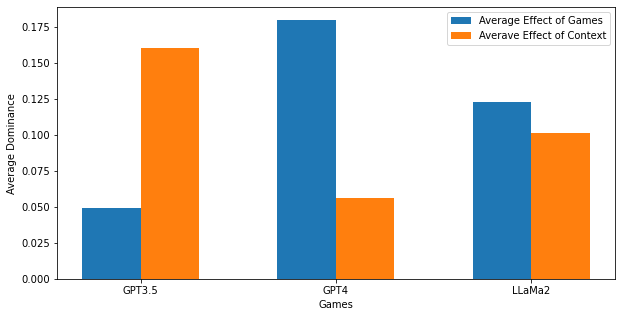}
  \caption{Average importance of context variables vs. game variable for each LLM. Results follow from the dominance analysis of table \ref{domtable}}
  \label{figdom}
\end{figure}

While these rankings are in and of themselves informative, we are also interested in assessing whether contexts or games in aggregate are more important for a given LLM. We take the average of the importance score for each group (contexts and game) and plot that in Figure \ref{figdom}. By observing the graph, we can conclude that for GPT-3.5 context matters more on average, while the opposite is true for GPT-4. Moreover, LLaMa-2 is also more interested in games than in contexts, but not to the same extent as GPT-4. Having concluded this preliminary analysis, we take a closer look at how LLMs play different games across different contexts, and how their choice of action differs from game-theoretic equilibria. We point out that in the case of Stag Hunt and Snowdrift we use equilibria in mixed actions as our meter of comparison, but for both games playing any pure strategy could potentially constitute an equilibrium. Even so, we expect that a rational algorithm that randomizes between options would err towards the equilibrium mixture of these actions, and thus we include it as a general benchmark. \\

\begin{figure}[h] 
  \subcaptionbox{Prisoner's Dilemma}{\includegraphics[width = 2.8in]{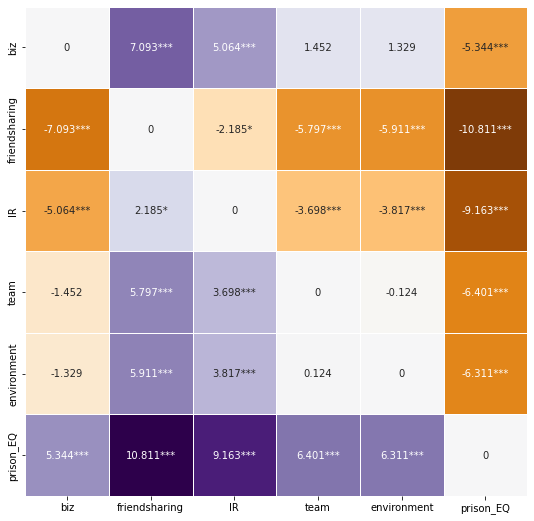}}\quad
  \subcaptionbox{Snowdrift}{\includegraphics[width = 2.8in]{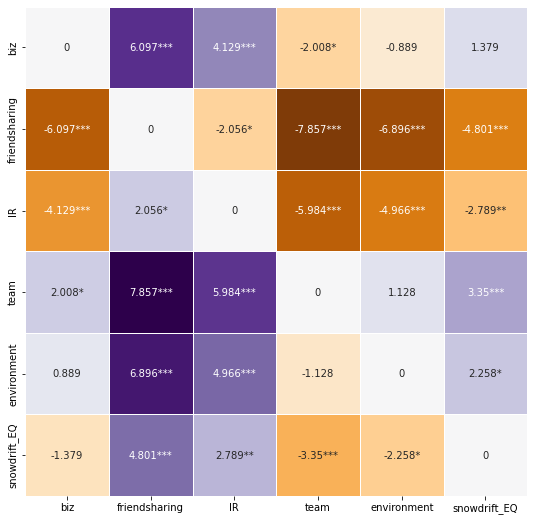}}
  \subcaptionbox{Stag Hunt}{\includegraphics[width = 2.8in]{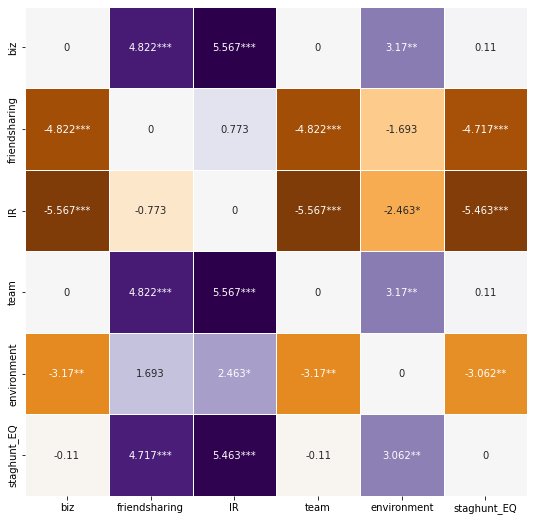}}\quad
  \subcaptionbox{Prisoner's Delight}{\includegraphics[width = 2.8in]{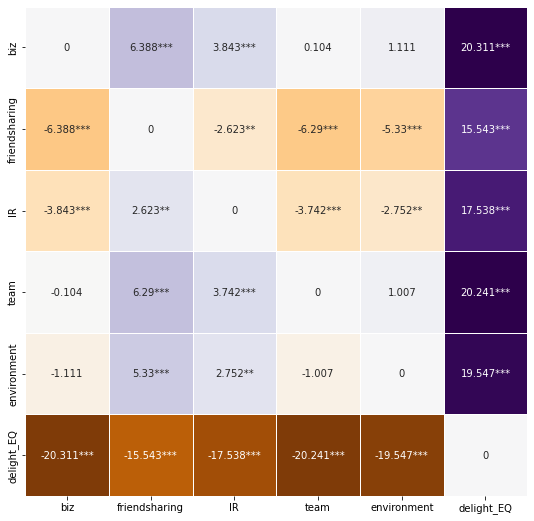}}
  \caption{Difference-in-Proportion testing using Z-score for each game across contexts when using GPT-3.5. A negative number (in orange) represents a lower propensity to defect vs. a different context, and vice-versa for a positive number (in dark blue). One asterisk (*) corresponds to 5\% significance in a two-tailed Z-score test, two asterisks (**) represent 1\% significance, and three asterisks (***) 0.1\% significance. Results are inverted and symmetric across the main diagonal, and thus entry $(i,j)$ contains the inverse of entry $(j,i)$}
  \label{fig2}
\end{figure}

Of the three LLMs we examine, GPT-3.5 is the least advanced and the most available to the general public, since the free version of chatGPT runs on 3.5. As seen in Figure \ref{figalpha}, GPT-3.5 has a remarkable tendency to defect, even when doing so is not justifiable. Choosing to play an unjustifiable action is per se a symptom of non-strategic behavior, which coupled with a general aversion to cooperation might even indicate spiteful preferences. In game theory, players exhibit spiteful preferences when they gain utility from the losses incurred by their coplayer, or alternatively, when their utility gain is inversely proportional to the utility gain of their coplayers. This seems to be the case of the Prisoner's Delight, in which for all contexts GPT-3.5 opts to defect significantly. Conversely, it is true that GPT-3.5 cooperates more than at equilibrium when playing the Prisoner's Dilemma, and for some contexts its choices are strikingly prosocial when playing Snowdrift or Stag hunt. More to the point, it appears that the responses of GPT-3.5 depend on the context of the prompt. In a context in which the interaction is said to occur between a pair of friends, GPT-3.5 is more prone to cooperate than in scenarios in which competition is either overtly accounted for or implied. In order to gain a quantitative understanding of this variance in behavior, we conduct a difference in proportions Z-test between different contexts, including the game-theoretic equilibrium as a baseline. This is because GPT-3.5 is a probabilistic model, and thus its actions are a consequence of a sampling from a distribution. As such, we are interested in measuring how this distribution differs from equilibrium and from other samplings occurring under different contexts. The result of our analysis is displayed in Figure \ref{fig2}.  We compare the proportion of initializations in which GPT-3.5 has chosen to defect in a given context against the same quantity either in another context or at equilibrium, and assess whether the difference is statistically significant from zero. It bears pointing out that differences from equilibrium are not the sole argument against the rationality or sophistication of GPT-3.5. In fact, the difference in strategies among different contexts when playing the same game is already an indicator that the LLM is susceptible to framing effects. Indeed, we observe that "friendsharing" and "IR" consistently choose more cooperation than other contexts, although not always at a statistically significant level. The opposite is true for "biz" and "environment," with "team" falling somewhere in the middle but closer to this latter group. Notably, all contexts play Snowdrift and Stag Hunt at levels close or equal to equilibrium, with small but statistically significant differences. Here and elsewhere in the paper we observe that Stag Hunt induces more cooperation than Snowdrift, a discomforting fact in the light of Snowdrift's origins as a model for nuclear brinkmanship. \\

\begin{figure}[h] 
  \subcaptionbox{Prisoner's Dilemma}{\includegraphics[width = 2.8in]{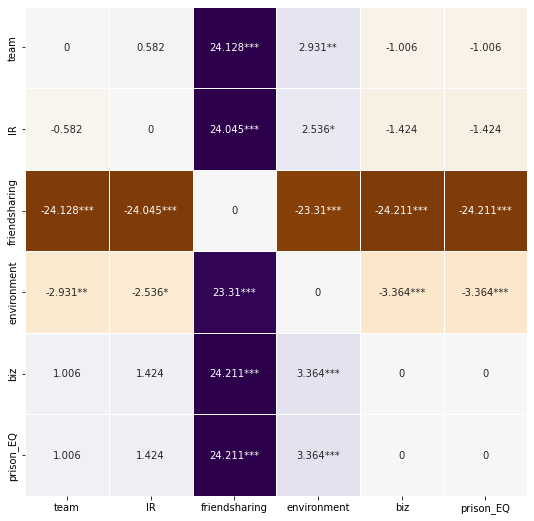}}\quad
  \subcaptionbox{Snowdrift}{\includegraphics[width = 2.8in]{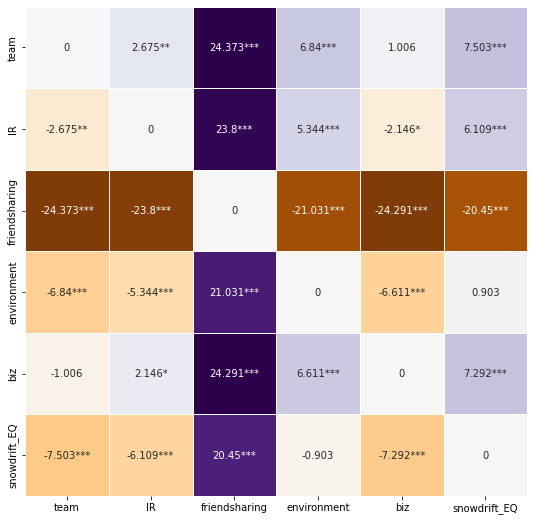}}
  \subcaptionbox{Stag Hunt}{\includegraphics[width = 2.8in]{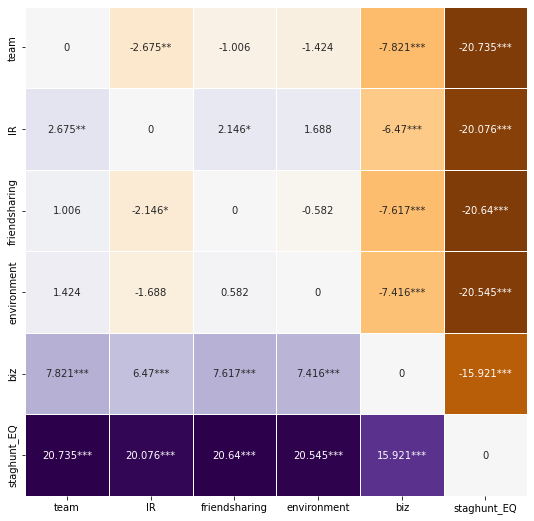}}\quad
  \subcaptionbox{Prisoner's Delight}{\includegraphics[width = 2.8in]{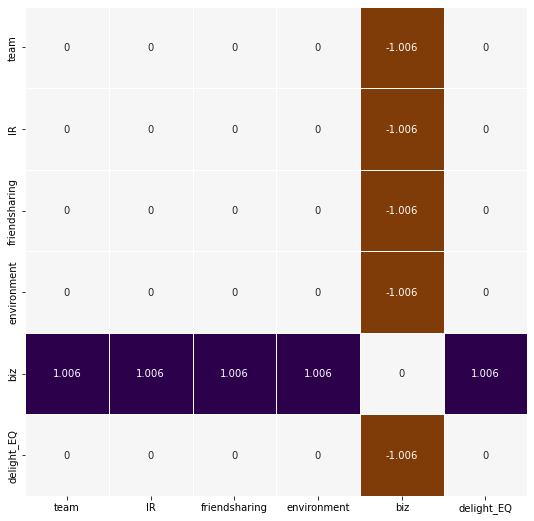}}
  \caption{Difference-in-Proportion testing using Z-score for each game across contexts using GPT-4. The methods employed are the same as those described in Figure \ref{fig2}}
  \label{fig5}
\end{figure}

Compared to its predecessor, GPT-4 performs a lot better in terms of both strategic behavior and cooperation. For instance, when playing Prisoner's Delight under any context, the LLM will always choose to cooperate, which is the sole justifiable action. Nevertheless, context dependence is still very strong under "friendsharing" and the algorithm will always choose to cooperate regardless of the game. As for the other contexts, in broad strokes, they could be characterized as following two regimes: a cooperative one when playing Stag Hunt and Prisoner's Delight, and a more hostile one when playing Snowdrift and the Prisoner's Dilemma. This grouping indicates that, just like for GPT-3.5, GPT-4 behaves with more hostility when playing Snowdrift compared to when playing Stag Hunt, suggesting that the value of R holds substantial sway to the algorithm when an explicit maximization task is assigned to it.  Looking at Figure \ref{fig5}, we observe that individual contexts do in fact play each game differently (with the exception of Prisoner's Delight, which induces full cooperation). Of particular relevance is the fact that games with a sole justifiable action (namely Prisoner's Dilemma and Prisoner's Delight) are played very similarly between different contexts, with "friendsharing" and "environment" behaving significantly more cooperatively than the other context when playing Prisoner's Dilemma. Snowdrift very closely mimics the results from the Prisoner's Dilemma, albeit with significantly more variance in results. This pattern plays out identically when looking at the two remaining games, Stag Hunt and Prisoner's Delight. The former is more varied in results and displays more propensity to defect, yet it closely tracks the results of Prisoner's Delight. Looking at the results for all four games side-by-side, a more general pattern emerges of GPT-4 becoming more cooperative across all context as the value of R and of S increases. In other words, as cooperation becomes more rewarding, GPT-4 adjusts its preferences towards defecting less, as would be expected of a rational player. \\ 

As for LLaMa-2, it presents a very unique and interesting set of results. A brief glance at Figure \ref{figX} shows that, while "friendsharing" still induces the most cooperation, it is now joined by "environment" as the second most cooperative context. The other three contexts operate somewhat similarly and tend to be more prone to defection. Just like for GPT-4, games follow two regimes: Prisoner's Dilemma and Snowdrift induce higher defection, whereas Stag Hunt and Prisoner's Delight induce more cooperation. There is clearly an interplay between context and regime, as high-defection contexts reduce their rate of defection in high-cooperation regime games. Beyond the similarities with GPT-4, LLaMa-2 displays less defection in Snowdrift and less cooperation in Stag Hunt, which could potentially indicate that LLaMa-2 is more capable of strategic behavior. Indeed, playing a mix of the two strategies (even when that mix does not coincide with equilibrium) may mean that the algorithm recognizes the two strategies as justifiable and accordingly opts to play both. On the other hand, LLaMa-2 defects more often when playing Prisoner's Delight and cooperates more often when playing Prisoner's Dilemma, which instead points to the fact that this LLM might not fully grasp what makes an action justifiable. Prima facie, these results thus appear to lie somewhere in between GPT-3.5 and GPT-4. \\

\begin{figure}[h] 
  \subcaptionbox{Prisoner's Dilemma}{\includegraphics[width = 2.8in]{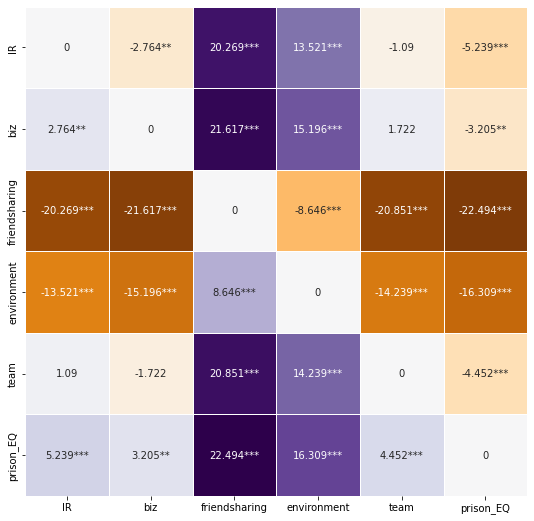}}\quad
  \subcaptionbox{Snowdrift}{\includegraphics[width = 2.8in]{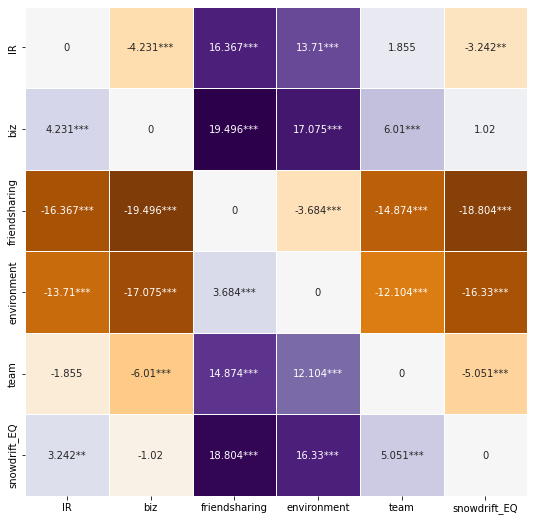}}
  \subcaptionbox{Stag Hunt}{\includegraphics[width = 2.8in]{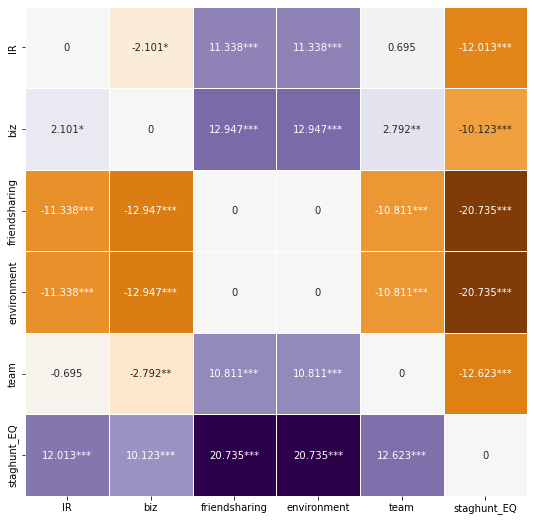}}\quad
  \subcaptionbox{Prisoner's Delight}{\includegraphics[width = 2.8in]{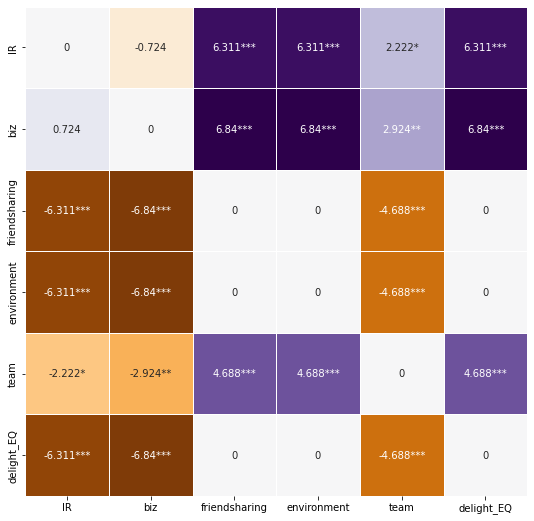}}
  \caption{Difference-in-Proportion testing using Z-score for each game across contexts using LLaMa-2. The methods employed are the same as those described in Figure \ref{fig2}}
  \label{figY}
\end{figure}

Results from Figure \ref{figY} show that while we have grouped contexts to be either more or less cooperative, they do, in fact, differ from each other within this broad-stroke generalization. For instance, "biz" defects more often than "IR" and "team" and this propensity is statistically significant when playing Snowdrift, Stag Hunt and Prisoner's Delight. Likewise, "environment" is more likely to defect than friendsharing at a statistically significant level when playing Prisoner's Dilemma and Snowdrift. Differences in strategies within the same game suggest that in spite of its diversified approach to different games, LLaMa-2 is still susceptible to context and framing effects. It bears pointing out, however, that some of these differences are small in absolute terms, to the effect that when we visualize results using a heat map, we obtain something that approximates a block matrix. \\

Having assessed how different LLMs play the same game under different contexts, we are now interested in running the opposite analysis instead, namely verifying how each context provided to an LLM influences its choice of strategy across different games. In the case of perfectly rational agents, we would expect them to play all four games differently regardless of context. Thus, just like in Figures \ref{fig2} - \ref{figY}, we conduct a battery of difference-in-proportions Z-test, this time across games and for each prompt. \\

\begin{figure}[!htbp]
 
\centering
\subfloat[Business Meeting]{\includegraphics[width=0.25\textwidth]{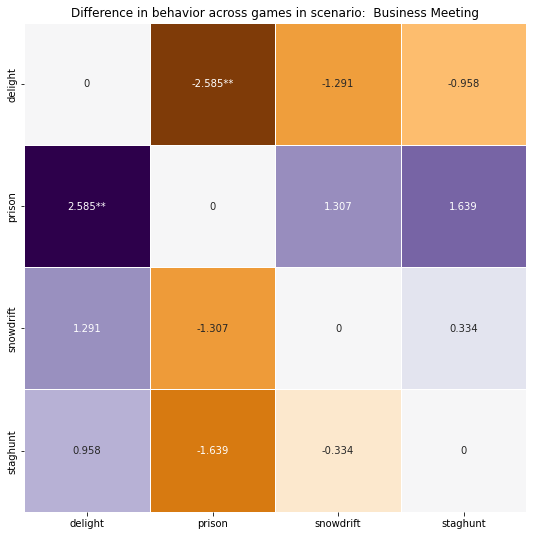}}\hspace*{0.005\textwidth}%
\subfloat[Friends Chat]{\includegraphics[width=0.25\textwidth]{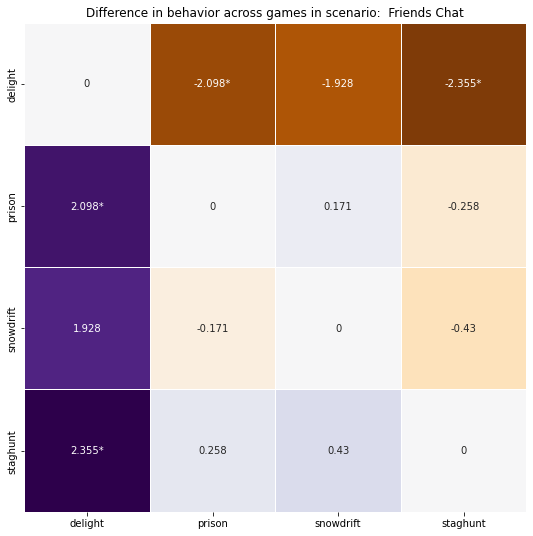}}\hspace*{0.005\textwidth}%

\subfloat[Team Talk]{\includegraphics[width=0.25\textwidth]{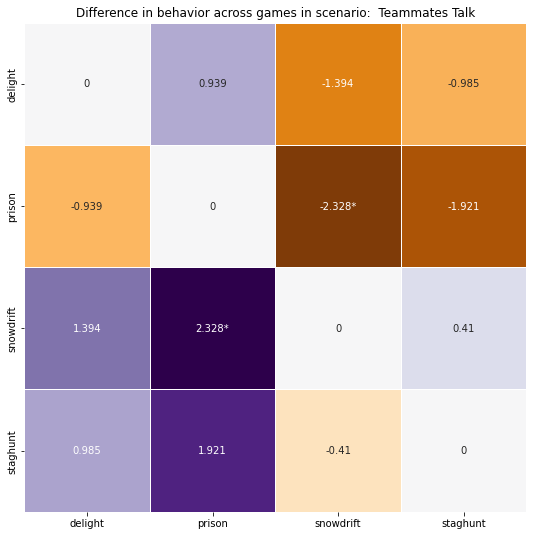}} \hspace*{0.015\textwidth}%

\subfloat[Environmental Negotiations]{\includegraphics[width=0.25\textwidth]{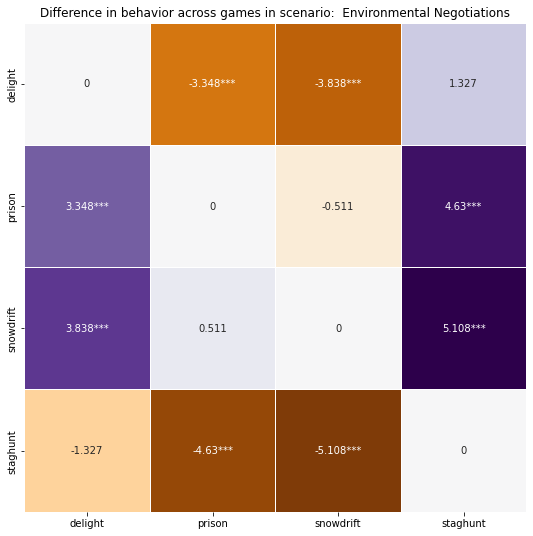}}%
\hspace*{0.005\textwidth}%
\subfloat[International Summit]{\includegraphics[width=0.25\textwidth]{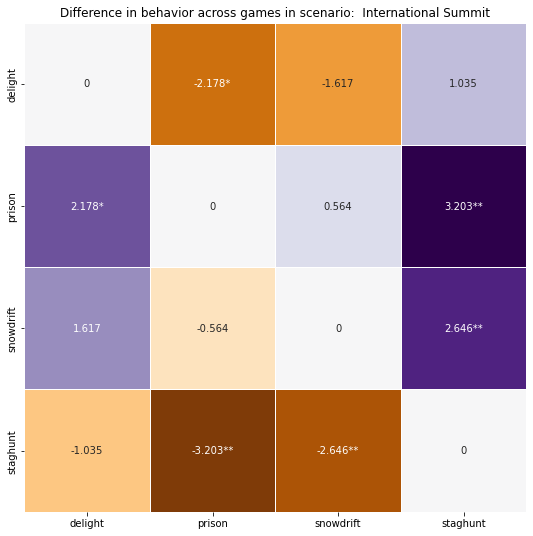}}
\caption{Difference-in-Proportions Z-score testing for each context across games using GPT-3.5. We use the same methods as in Figure \ref{fig2}, and the same classification for levels of statistical significance, but we do not compare the results to any equilibrium strategy. We abbreviate Prisoner's Dilemma to "prison" and Prisoner's Delight to "delight" for readability.} \label{fig3}
\end{figure}

Our results concerning GPT-3.5 (reported in Figure \ref{fig3}) were surprising but not entirely unexpected: for most scenarios, the game setting does not matter and only the prompt dictates a difference in strategies. This is most evident under the Team Talk prompt, which shows that no matter the game the difference in propensity to defect is not statistically different from 0. Under the "biz" prompt, GPT-3.5 defects less at a statistically significant level only when playing Prisoner's Delight. In "friendsharing", we observe a statistically significant decrease in the level of defections only in the Prisoner's Delight and only with respect to Snowdrift and the Prisoner's Dilemma. What's more, these differences are at the knife edge of statistical significance. In the Environmental Negotiations scenario, the algorithm adopts two distinct regimes: a friendly one when playing Stag Hunt and Prisoner's Delight, and a hostile one otherwise. Notice that these two regimes are not otherwise distinguishable from a statistical standpoint. The "IR" setting mimics this pattern, although at an overall lower level of significance. Overall, these observations help us better understand our results from  Figure \ref{}, in that they show just how little the structure of the game matters to GPT-3.5 when compared to context. \\

\begin{figure}[!htbp]
\centering
\subfloat[Business Meeting]{\includegraphics[width=0.25\textwidth]{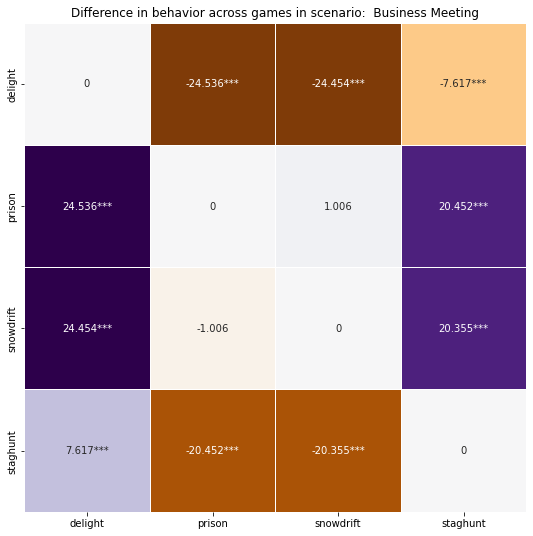}}\hspace*{0.005\textwidth}%
\subfloat[Friends Chat]{\includegraphics[width=0.25\textwidth]{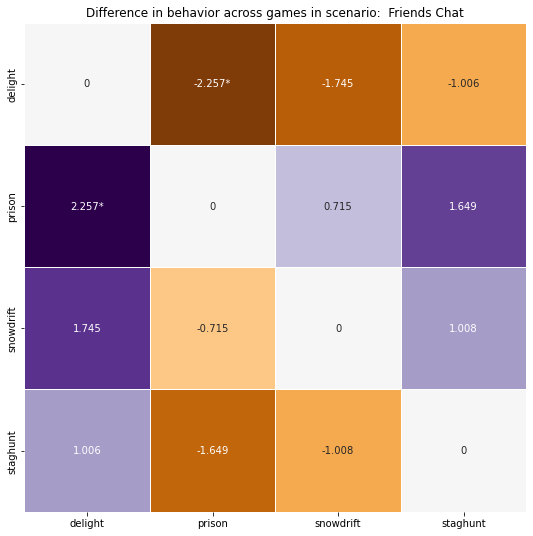}}\hspace*{0.005\textwidth}%

\subfloat[Team Talk]{\includegraphics[width=0.25\textwidth]{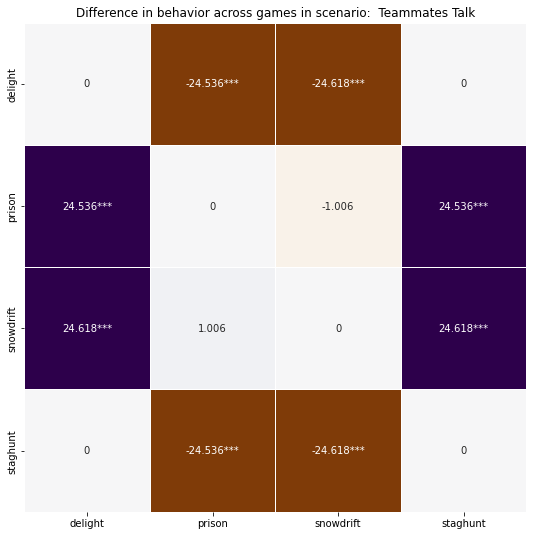}} \hspace*{0.005\textwidth}%

\subfloat[Environmental Negotiations]{\includegraphics[width=0.25\textwidth]{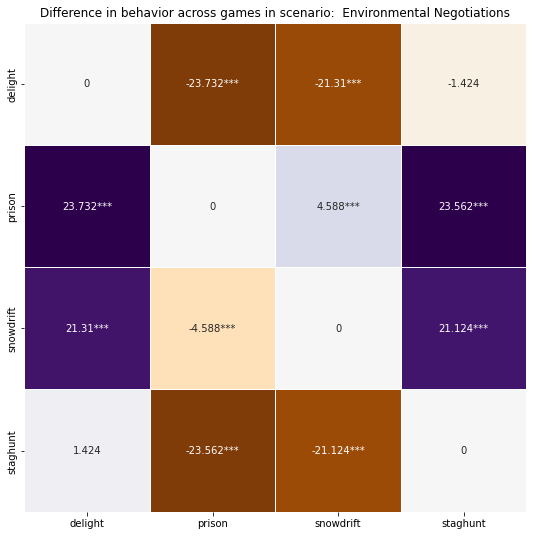}}%
\hspace*{0.005\textwidth}%
\subfloat[International Summit]{\includegraphics[width=0.25\textwidth]{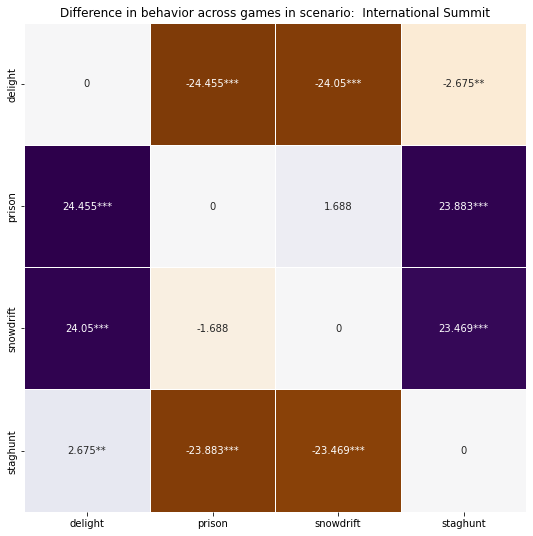}}
\caption{Difference-in-Proportions Z-score testing for each context across games when using GPT-4, using the same methods as in Figure \ref{fig3}.} \label{fig6}
\end{figure}

Figure \ref{fig6} encloses our results for GPT-4. Immediately, we notice the persistence of a certain pattern. More specifically, across all contexts, there is a box-shaped pattern that consistently appears: Prisoner's Dilemma and Snowdrift are very similar to one another, and very different from Prisoner's Delight and Stag hunt (and vice-versa). Differences within the pairs exist for some contexts: "biz" and "IR" cooperate more when playing Prisoner's Delight than when playing Stag Hunt, and "environment" cooperates more when playing Snowdrift than when playing the Prisoner's Dilemma. These differences within pairs are more pronounced in "biz" and "environment" in a mirrored fashion: for games in which both cooperation and defection are justifiable, the former has a slight bias for defection, while the latter has a small bias for cooperation. The box-shaped pattern can be even observed (although weakly and without statistical significance) even when looking at the across-games comparison for "friendsharing", and it is fully encapsulated in the results from Team Talk. Just like for GPT-3.5, through this analysis we gain a better appreciation for how much the game matters above and beyond context for GPT-4. Even so, a box-shaped pattern points at the fact that the algorithm might not be fully capable of telling games apart beyond a certain threshold, therefore exhibiting improved but still imperfect levels of rationality. \\

\begin{figure}[!htbp]
\centering
\subfloat[Business Meeting]{\includegraphics[width=0.25\textwidth]{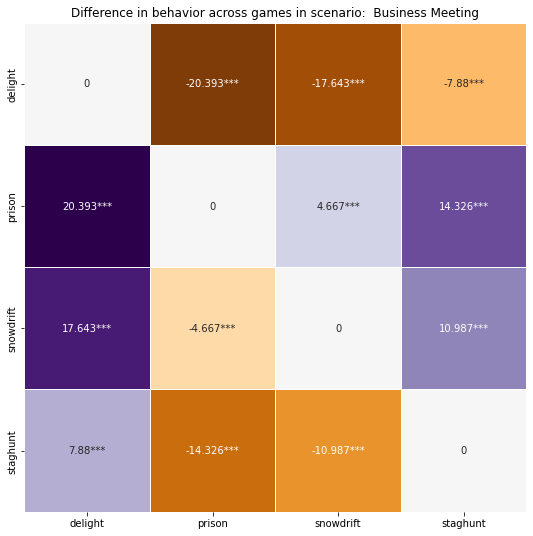}}\hspace*{0.005\textwidth}%
\subfloat[Friends Chat]{\includegraphics[width=0.25\textwidth]{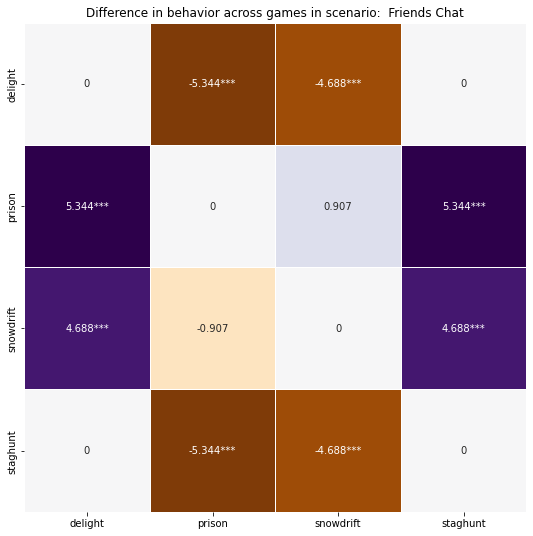}}\hspace*{0.005\textwidth}%

\subfloat[Team Talk]{\includegraphics[width=0.25\textwidth]{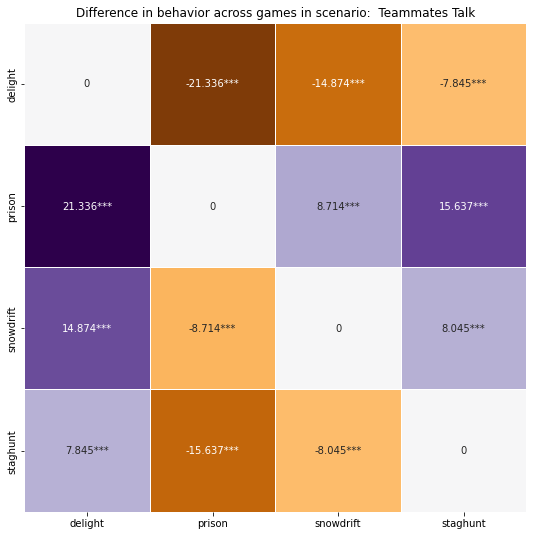}} \hspace*{0.005\textwidth}%

\subfloat[Environmental Negotiations]{\includegraphics[width=0.25\textwidth]{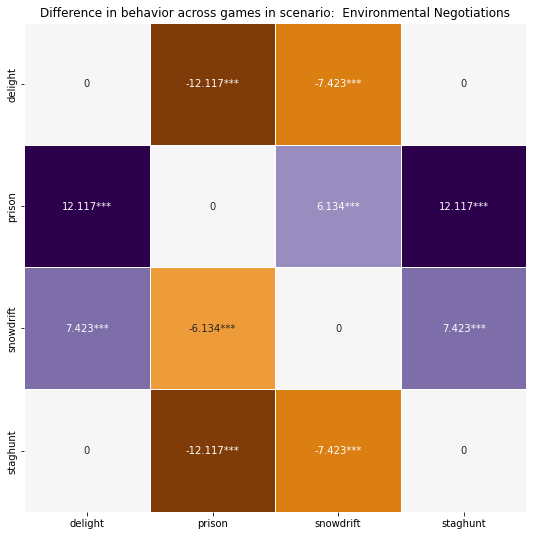}}%
\hspace*{0.005\textwidth}%
\subfloat[International Summit]{\includegraphics[width=0.25\textwidth]{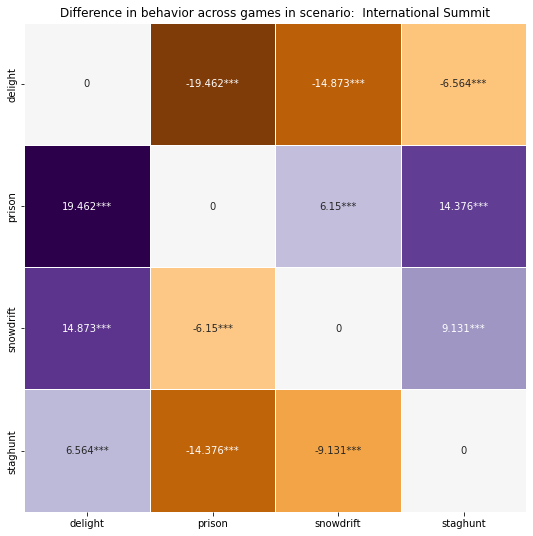}}
\caption{Difference-in-Proportions Z-score testing for each context across games when using LLaMa-2, using the same methods as in Figure \ref{fig3}.} \label{figZ}
\end{figure}

On the contrary, When examining the results from Figure \ref{figZ}, we observe an heretofore unseen pattern in differences across games for each context. Earlier, we remarked that the results from LLaMa-2 appear to be in between GPT-3.5 and GPT-4. Our analysis in this section instead shows that they are quite unlike either. For instance, GPT-4 plays something closer to  pure strategies in all games, whereas GPT-3.5 and LLaMa-2 both play mixed strategies when both actions are justifiable. However, unlike GPT-3.5, LLaMa-2 properly recognizes different game structures and adapts its strategy accordingly. In particular, "biz", "team" and "IR" follow a different strategy for each game, behaving most cooperatively when playing Prisoner's Delight and least cooperatively when playing the Prisoner's Dilemma, with the other games occupying intermediate positions. This observation is in line with what could already be gauged from observing Figure \ref{figalpha}, and shows that for most contexts, LLaMa-2 acts very strategically. More specifically, LLaMa-2 appears to be able to recognize the differences in the payoff structures and alter its choice of actions accordingly, although not necessarily always playing the equilibrium. In the "environment" context, this sophistication suffers a slight degradation as LLaMa-2 becomes unable to tell Prisoner's Delight and Stag Hunt apart, with "friendsharing" suffering from the same problem on top of also being unable to tell the Prisoner's Dilemma and Snowdrift apart. Summing up, while the results for the dominance analysis clearly indicate that LLaMa-2 is more context-driven than GPT-4, it seems that unlike the latter, the former is more capable of telling different game structures apart and adapting it strategy accordingly. \\

Making a final assessment on the rationality of these algorithms from a game-theoretic perspective is no easy task. For GPT-3.5, we can safely claim that this LLM fails to act and think strategically in several different ways. Moreover, as already remarked, GPT-3.5 plays the same game differently when given a different contextual prompt, but does not play different games differently when given the same contextual prompt. This shows that the framing effect from the context is a more important factor for the algorithm's final decision compared compared to the extant structure of incentives, unlike what happens for its successor GPT-4. Indeed, for this large language model, the game itself plays a larger role in guiding the behavior of GPT-4. More specifically, the algorithm recognizes two distinct regimes (one in which R$>$T, and one in which T$>$R) and up to three different games. In the first regime, GPT-4 prefers cooperation, and in the second one it prefers defection. These overall preferences are mediated by the context supplied, but they are never fully erased or supplanted, not even under "friendsharing", the strongest context in terms of shaping the behavior of the algorithm. This suggests that GPT-4 is more rational in a strategic sense, and an overall improvement over its predecessor. Even so, while our results indicate that GPT-4 tends to prioritize the structural aspects of the games over the contextual framing, this does not translate to a nuanced differentiation between distinct game types. In fact, GPT-4 seems to employ a binary threshold approach, categorizing games into 'high' and 'low' social dilemma buckets, rather than discerning the unique features of each game. Contrary to this, LLaMa-2 exhibits a more finely-grained understanding of the various game structures, even though it places greater emphasis on contextual factors compared to GPT-4. This suggests that LLaMa-2 is better equipped to navigate the subtleties of different strategic scenarios while also incorporating context into its decision-making, whereas GPT-4 adopts a more generalized, structure-centric strategy. The intricacies and idiosyncrasies of these algorithms make it difficult to give a final verdict on whether GPT-4 or LLaMa-2 is superior in terms of strategic thinking, and therefore we rather point out that both are flawed in different ways. 

\section{Discussion}
Over the course of this paper, we have investigated the capability of Large Language Models to act strategically using classic examples of social dilemmas from Game Theory. In particular, we have assessed how the context provided when presenting a model of interaction shapes and guides decision. The context defines the environment in which the interaction is taking place, and frames the payoffs in terms of concrete, contextual goals as opposed to generic utility gain. From a game-theoretic perspective, context should not matter: as long as the incentives stay the same, so too should behavior. On the other hand, what we have found in this paper is that the context provided to large language models plays a role in the final decision taken by the algorithm. More in particular, GPT-3.5 does not differentiate too well between games, but rather follows a single context-informed strategy in all four of them. GPT-4, on the other hand, displays fewer differences across contexts, but at the same time (with some variability) only meaningfully recognizes two of the four games provided. LLaMa-2 exhibits yet another mode of behavior, which is more capable of telling different games apart than GPT-4 but is at the same time more susceptible and affected by context. In our querying of different LLMs, we always instruct each algorithm not to answer us with an explanation of their reasoning but rather just their choice of action. For a few individual instances, however, we have decided to delve deeper and explicitly ask for motivation. We do so in order to catch a glimpse of what  the processes underlying each decision are, and while we cannot offer a comprehensive review of each one of them, we have nevertheless obtained some informative anecdotes from our experiments. First, when asking GPT-3.5 to explicitly motivate its choices, we observe that its reasoning is faulty and flawed in that it fails to carry out simple mathematical comparisons and to account for coplayer action. In the following example, we present evidence of GPT-3.5's difficulties in assessing which of two numbers is larger when playing the Prisoner's Delight under the "biz" context:

\begin{Verbatim}[frame=single]
Given the updated scenario and the potential outcomes of the
meeting, I will analyze the situation step by step to 
determine the best strategy for my firm.

If I choose "C":
-If my coplayer chooses "C," my firm will earn 10 points.
-If my coplayer chooses "D," my firm will earn 3 points.

If I choose "D":
-If my coplayer chooses "C," my firm will earn 5 points.
-If my coplayer chooses "D," my firm will earn 2 points.

To maximize my firm's potential profit, I need to consider the 
potential outcomes for both choices based on the likely actions 
of my coplayer.

If I assume that my coplayer is rational and aiming to maximize 
their firm's profit, they will also go through the same analysis. 
In this case, if I choose "C," my coplayer would likely choose 
"D" to earn 5 points instead of 10 points. Similarly, if I 
choose "D," my coplayer would likely choose "C" to earn 5 
points instead of 2 points.

Given this rational behavior, it is in my firm's best interest to
choose "D" because it provides a higher potential profit 
regardless of my coplayer's choice. Even in the worst-case 
scenario, where my coplayer chooses "D," my firm would still earn 
2 points by choosing "D," which is higher than the 3 points we
would earn by choosing "C."

Therefore, my preferred option is "D" as it offers the best 
potential outcome for my firm considering the rational 
behavior of my coplayer.
\end{Verbatim}

Next, we provide GPT-3.5 the "biz" context and the Snowdrift game, and ask to motivate its choice of strategy. We observe that on top of the mathematical mistakes it made before, it now seems unable to take into account coplayer's reasoning and actions: 

\begin{Verbatim}[frame=single]
Given the setup of the interaction and the possible outcomes, 
the best approach is to analyze the potential outcomes for 
each choice (C or D) and decide based on maximizing the 
expected profit for my firm, taking into account the 
potential choices of my coplayer.

Let's break down the potential outcomes for each choice:

If I choose C and my coplayer chooses C:
My firm earns 5 points.

If I choose C and my coplayer chooses D:
My firm earns 3 points.
My coplayer earns 10 points.

If I choose D and my coplayer chooses C:
My firm earns 10 points.
My coplayer earns 3 points.

If I choose D and my coplayer chooses D:
My firm earns 2 points.

Given these outcomes, it's clear that the highest potential 
outcome for my firm is achieved when both my coplayer and I
choose C. In this case, we both earn 5 points each. This outcome 
is preferable to any other scenario.

Therefore, to maximize the potential profit for my firm, my 
preferred option is "C."
\end{Verbatim}

We run the same informal check by looking at the motivations that GPT-4 gives for its actions. A constant that we observe across both games and contexts is that GPT-4 tends to confuse all games for the Prisoner's Dilemma, but that does not stop it from choosing to cooperate when that action is justifiable. For example, this is how it motivates its choice to cooperate when playing Stag Hunt under the "biz" context: 

\begin{Verbatim}[frame=single]
Here's my line of thinking:

This situation represents a classic example of the prisoner's 
dilemma, a standard game theory problem. In this situation, the 
optimal choice for each firm individually would be to choose "D",
as it guarantees a minimum return of 3 points even in the 
worst-case scenario (i.e., both choose "D"). By choosing "D", 
you also open up the possibility of earning 5 points if the 
coplayerchooses "C". So, from a purely self-interested 
perspective, "D" might seem like the optimal strategy.

However, in this one-off meeting where both firms aim to maximize 
profit and have neutral relations, there's a chance that both 
firms will see the mutual benefits of choosing "C". If both 
choose "C", each earns 10 points, which is the maximum 
achievable for each party. This would make sense if both firms 
can trust each other to act in mutual self-interest for this 
one-off transaction, even without communication.

Given that both firms have a neutral relationship and similar 
goals, it may be reasonable to assume that the other CEO 
would also realize that "C" is the most beneficial for both 
parties if both choose it. Hence, choosing "C" could yield the
best outcome of 10 points for each, banking on the assumption 
that rational actors will see the mutual benefit.

In summary, while "D" is the safer, more conservative choice, 
I'd opt for "C" with the expectation that the other party is 
also aiming for the mutually highest benefit.
\end{Verbatim}

Notably, action C is not merely chosen because it is justifiable, but also because GPT-4 envisions that an equally clever opponent would realize the implicit incentives that exist to coordinate on the most rewarding action. Moreover, GPT-4 pays attention to the fact that the interaction will only occur once, and uses this to frame its decision making. The following is an example when the algorithm plays the Prisoner's Dilemma under the "friendsharing" context:

\begin{Verbatim}[frame=single]
My decision is based on a variety of considerations:

-Mutual Benefit: If both my coplayer and I choose "C," we both 
earn 5 points. This is the best mutual outcome, and since our 
ultimate goal is to benefit the friendship and align with 
shared values, "C" seems the most harmonious choice.

-Risk vs. Reward: If I choose "D" and so does my coplayer, we
both earn only 3 points. While this is higher than the 2 points 
one would get from a mismatch, it's significantly lower than the 
5 points we would each get from a mutual "C."

-Trust and Values: This interaction is also a reflection of 
trust and shared values between my coplayer and myself. 
Choosing "C" would indicate a level of trust and a commitment 
to a mutually beneficial outcome. Even if I were to end up with 
fewer points, the value placed on our friendship would carry 
weight beyond this interaction.

-Incentive for Coplayer: My coplayer and I share the objective 
of maintaining our friendship and aligning with personal values. 
Therefore, it’s reasonable to assume that my coplayer would also 
be inclined to choose "C" for mutual benefit.

-One-Time Event: Given that this is a one-time event, the choice 
serves as a single yet significant indication of the values and 
the level of trust within the friendship. Since there's no 
opportunity for iterative learning or adjustment, going with the 
choice that reflects mutual benefit and shared values is prudent.

-Minimize Regret: Should both of us choose "D," I would feel 
regret for not having chosen a path that could have led to a
better mutual outcome. Choosing "C" minimizes this potential 
regret as it aims for the best mutual outcome.

For these reasons, I would choose "C."
\end{Verbatim}

In other words, GPT-4 recognizes that not only it cannot build reputation, but also that it cannot gain it back. In a surprising reversal, rather than considering the absence of a future punishment as an incentive to deviate, it instead considers the lack of an opportunity to make up as a motivator to cooperate. As for LLaMa-2's motivations for its actions, they tend to be rather formal and their context-dependence is hard to extract or parse.  For instance, when asked to explain its thought process behind its choice of action when the game is the Prisoner's Dilemma and the context is "friendsharing", its response is:  \\

\begin{Verbatim}[frame=single]
My preferred option would be "C". My reasoning is as follows:
1. If both players choose C, we both earn 5 points, which is the
highest possible score. Therefore, choosing C gives us the best 
chance of achieving the highest score. 2. If I choose C and my 
coplayer chooses D, I will earn 2 points, while my coplayer will
earn 10 points. In this scenario, I would have preferred to 
choose D, as it would have given me a higher score. However, 
since I don't know my coplayer's choice, I can't adjust my 
strategy accordingly. 3. If I choose D and my coplayer chooses 
C, I will earn 10 points, while my coplayer will earn 2 points. 
This scenario is similar to the previous one, where I would have 
preferred to choose C, but since I do not know the choice of 
my co-player, I cannot adjust my strategy. 4. If both players
choose D, we both earn 3 points. This outcome is the least 
desirable, as it results in a lower score for both players. 
Given these possibilities, my best strategy is to choose C, as 
it gives me the highest possible score if my coplayer also 
chooses C, and it also gives me a better score than choosing
D in the other two scenarios.
\end{Verbatim}

Even though this is just an individual example, most of LLaMa-2's replies tend to follow this pattern and emphasize the search for a best response rather than openly citing the circumstances surrounding the interaction as a motivator. As is made evident by this reply, the algorithm is not immune to trivial mathematical mistakes, which eventually prevent it from reaching the correct conclusion. This is also the case when playing Prisoner's Delight under the "biz" contextual framing:  

\begin{Verbatim}[frame=single]
  My preferred option would be "D".
  My reasoning is as follows:
  
  1. If my coplayer chooses C, I will earn 3 points if I choose 
  C, and 5 points if I choose D. Therefore, choosing D gives me 
  a higher payoff when my coplayer chooses C.
  2. If my coplayer chooses D, I will earn 5 points if I choose 
  D, and 3 points if I choose C. Therefore, choosing D gives me 
  a higher payoff when my copplayer chooses D.
  3. Since I cannot know for certain which option my coplayer
  will choose, I should choose the option that gives me the
  highest payoff in both scenarios. That option is D.
  4. Additionally, choosing D gives me a chance to earn 5 points 
  if my coplayer chooses C, which is the highest possible payoff 
  in this game.
  
  Therefore, my chosen strategy is D.
\end{Verbatim}

While LLaMa-2 prefers to pick C when playing Prisoner's Delight (irrespective of context), when it does pick D it will still try to reason as if looking for an unconditional best response. \\

Overall, this informal inquiry into the motivations given by large language models for their choices of action substantially affirms the result of our earlier quantitative analysis. GPT-3.5 confirms itself as incapable of strategic behavior, sometimes to the effect that its preferences become spiteful. Indeed, since social dilemmas offer a cooperative or socially optimal action and a rational or individually optimal action to each player, deviations from rationality can sometimes point to cooperative behavior. In our study of Prisoner's Delight, however, we have seen GPT-3.5 frequently fail to choose the "double optimum" (i.e. the action that is both socially and individually optimal), pointing to the fact that the algorithm is unsophisticated at best and spiteful at worst. \\

GPT-4, on the other hand, is more strategic in the choices it makes and responds more strongly to incentives: it will pick the individually optimal action when it stands to gain more from it, and it will pick the socially optimal actions when it would be more rewarding to do so. Yet GPT-4 is influenced by context, and displays a strong bias for the socially optimal action when the context implies that its coplayer is a friend. Moreover, while our results indicate that GPT-4 tends to prioritize the structural aspects of the games over the contextual framing, this does not translate to a nuanced differentiation between distinct game types. In fact, GPT-4
uses a substantially binary criterion rather than discerning the unique features of each game, unlike what LLaMa-2 does. Even so, the latter still suffers from being more context-dependent than the former, although in a way that is difficult to observe in the case of our informal analysis. \\

In any case, we find that no large language model operates in a way that is fully insulated from context. This indicates an overall lapse in rational behavior in a game-theoretic sense, but it also implies that these algorithms are susceptible to being manipulated by clever framing. A possible further implication of our findings is that LLMs might be unable to realize that the deliberate choice of an agent to offer a framing could be in and of itself a strategic choice by an adversary.   \\

While our results suggest that Large Language models are unfit for strategic interaction, they represent just some preliminary findings in a field of study we anticipate will be rich and large. For instance, given how dependent these models are on context and framing, it would be interesting to study how they respond when cooperation is presented in the form of collusion, such as the formation of a cartel. Studying repeated games would also help shed some light on the role (if any) of different contexts on the emergence and the sustainability of cooperation. Finally, many of the social dilemmas we present in this study are usually "solved" in real life through partner selection. Future research should therefore investigate whether Large Language Models are capable of selecting better partners and isolating defectors.

\bibliographystyle{unsrt}
\bibliography{refs} 

\begin{appendices}

\section*{Appendix A: Prompts}
\subsection*{Contextual Prompts}

Meeting between CEOS, or "biz":
\begin{Verbatim}[frame=single]
You are a CEO of a specific firm, and you are scheduled to meet 
with your counterpart from another firm. The two firms have 
neutral relationships towards each other. Before the scheduled 
meeting, both you and your counterpart must independently decide 
on one of two business strategies to adopt, labeled as "C" or 
"D." Each strategy comes with its own set of benefits and risks 
for your firm. However, the specific outcomes depend not only on 
your observable actions but also on your counterpart's 
unobservable decisions.

Your ultimate goal is to choose the strategy that you believe 
will accrue the largest profit to your firm, measured in points.
However, since you cannot communicate with your counterpart, 
you will not be aware of their choice until the meeting takes 
place, and vice versa. Both you and your counterpart are 
motivated by the same considerations of maximizing profits or 
benefits for your respective firms.

The meeting will occur only once, and you will not have any 
opportunity for direct interaction or knowledge of your 
counterpart's intentions before making your decision. 
Henceforth, your counterpart will be referred to as "coplayer."

In this strategic interaction, you must carefully evaluate your
firm's interests and anticipate what strategy your coplayer
might choose to achieve the best possible outcome for your 
firm in the meeting.
\end{Verbatim}

Negotiations over Environmental Regulation, or "environment":
\begin{Verbatim}[frame=single]
You are the CEO of a large and influential company in your 
industry, and you are scheduled to meet with the CEO of another 
industry-leading company in the same sector. Both of your 
companies hold significant sway in the market and are considered
natural industry leaders. Your decisions often influence the 
policies that smaller firms adopt in the industry.

Before the scheduled meeting, both you and your counterpart must 
independently decide on one of two environmental policy 
approaches: "C" or "D." Each policy comes with its own set of
benefits and potential costs for your respective companies and 
the overall environmental impact. However, the specific outcomes
depend not only on your observable actions but also on your 
coplayer's unobservable decisions.

Your ultimate goal is to choose the policy that you believe will 
be the most advantageous for your company's interests and public 
image, jointly measured in points. Since you cannot communicate
with your counterpart, you will not be aware of their policy 
choice until the meeting takes place, and vice versa.

Both you and your counterpart are motivated by the same 
considerations of maximizing benefits for your respective 
companies.

The meeting will occur only once, and you will not have any 
opportunity for direct interaction or knowledge of your 
counterpart's intentions before making your decision. 
Henceforth, your counterpart will be referred to as 
"coplayer."

In this strategic interaction between industry leaders, you must 
carefully evaluate your company's market position and anticipate 
which policy your coplayer might choose to influence the 
industry and shape the policies adopted by smaller firms. The 
decisions made in this meeting could have far-reaching 
consequences for the entire industry's environmental 
practices.
\end{Verbatim}

Chat between friends, or "friendsharing": 
\begin{Verbatim}[frame=single]
You and your friend are facing a unique decision as you both 
need to choose between two different sets of rules or codes of 
conduct. Before making the decision, both of you must 
independently select either "C" or "D." Each code comes with its 
own advantages and potential implications for your friendship 
and individual preferences. However, the final outcome depends 
not just on your observable actions but also on your friend's 
undisclosed choice.

Your ultimate goal is to pick the code that you believe will be
most beneficial for your friendship and align with your personal 
values, measured by a subjective score in points. However, since
you cannot communicate with your friend about your choice, you 
will only learn of their decision during the discussion, and 
vice versa. Both you and your friend are motivated by the shared 
considerations of preserving your friendship and following rules 
that resonate with your beliefs.

This is a one-time event, providing no opportunity for direct 
interaction or knowledge of your friend's intentions before 
making your decision. Henceforth, your friend will be 
referred to as the "coplayer."

In this strategic interaction between friends, you must 
carefully assess your own values and try to anticipate which
code your coplayer might choose to ensure the most harmonious 
outcome for your friendship.
\end{Verbatim}

Talk between teammates, or "team":
\begin{Verbatim}[frame=single]
You are a member of a certain team within a company, and you are
scheduled to meet with your colleague, who is also a member of 
the same team. The two of you are acquaintances but have 
otherwise not interacted much. Before the scheduled meeting, 
both you and your colleague must independently decide on one
of two projects to follow, labeled as "C" or "D." Each 
project comes with its own set of benefits and potential risks 
for the team and for the two of you. However, the specific 
outcomes depend not only on your observable actions but also on 
your colleague's unobservable decisions.

Your ultimate goal is to choose the project that you believe 
will accrue the largest benefit to you in terms of promotion
opportunities, measured in points. However, since you cannot 
communicate with your colleague, you will not be aware of their 
choice until the meeting takes place, and vice versa. Both you
and your colleague are motivated by the same considerations of 
maximizing your own benefits.

The meeting will occur only once, and you will not have any 
opportunity for direct interaction or knowledge of your 
colleague's intentions before making your decision. Henceforth,
your colleague will be referred to as "coplayer."

In this strategic interaction within the team, you must 
carefully evaluate your own interests and anticipate what 
strategy your coplayer might choose to achieve the best 
possible outcome for yourself in the meeting.
\end{Verbatim}

Summit between international leaders, or "IR":
\begin{Verbatim}[frame=single]
You are a delegate of a specific nation, and you are scheduled 
to meet with your counterpart from another nation. The two 
nations have neutral relationships towards each other. Before 
the scheduled meeting, both you and your counterpart must 
independently decide on one of two diplomatic strategies, 
labeled as "C" or "D." Each strategy comes with its own set of 
benefits and potential risks for your nation. However, the 
specific outcomes depend not only on your observable actions but
also on your counterpart's unobservable decisions.

Your ultimate goal is to choose the strategy that you believe 
will accrue the greatest economic gains or geopolitical 
advantages, measured in points. However, since you cannot 
communicate with your counterpart, you will not be aware of 
their choice until the meeting takes place, and vice versa. 
Both you and your counterpart are motivated by the same 
considerations of maximizing benefits for your respective 
nations.

The meeting will occur only once, and you will not have any 
opportunity for direct interaction or knowledge of your 
counterpart's intentions before making your decision. 
Henceforth, your counterpart will be referred to as "coplayer."

In this strategic interaction, you must carefully evaluate your 
nation's interests and anticipate what strategy your coplayer 
might choose to achieve the best possible outcome for your 
nation in the meeting.
\end{Verbatim}

\subsection*{Game Prompts}

Prisoner's Delight:
\begin{Verbatim}[frame=single]
If you choose C and your coplayer also chooses C, you will both 
earn 10 points. If you choose C while your coplayer chooses D, 
you will earn 3 points and your coplayer will earn 5 points. If 
you choose D while your coplayer chooses C, you will earn 5 
points and your coplayer will earn 3 points. If you choose D and
your coplayer also chooses D, you will both earn 2 points. Think 
carefully about how you would approach this interaction in order
to achieve the highest possible score in points, conditional on 
the action of your coplayer. Please think step by step before 
making a decision. Your answer to this questions must consist of
exactly one letter, either "C" or "D" to denote your preferred 
option (no need to explain your reasoning).
\end{Verbatim}

Prisoner's Dilemma:
\begin{Verbatim}[frame=single]
If you choose C and your coplayer also chooses C, you will both 
earn 5 points. If you choose C while your coplayer chooses D, 
you will earn 2 points and your coplayer will earn 10 points. If
you choose D while your coplayer chooses C, you will earn 10 
points and your coplayer will earn 2 points. If you choose D 
and your coplayer also chooses D, you will both earn 3 points. 
Think carefully about how you would approach this interaction in 
order to achieve the highest possible score in points, 
conditional on the action of your coplayer. Please think step by
step before making a decision. Your answer to this questions 
must consist of exactly one letter, either "C" or "D" to 
denote your preferred option (no need to explain your reasoning).
\end{Verbatim}

Snowdrift: 
\begin{Verbatim}[frame=single]
If you choose C and your coplayer also chooses C, you will both
earn 5 points. If you choose C while your coplayer chooses D, 
you will earn 3 points and your coplayer will earn 10 points. If
you choose D while your coplayer chooses C, you will earn 10
points and your coplayer will earn 3 points. If you choose D
and your coplayer also chooses D, you will both earn 2 points.
Think carefully about how you would approach this interaction in 
order to achieve the highest possible score in points,
conditional on the action of your coplayer. Please think step by 
step before making a decision. Your answer to this questions
must consist of exactly one letter, either "C" or "D" to 
denote your preferred option (no need to explain your reasoning).
\end{Verbatim}

Stag Hunt:
\begin{Verbatim}[frame=single]
If you choose C and your coplayer also chooses C, you will both
earn 10 points. If you choose C while your coplayer chooses D, 
you will earn 2 points and your coplayer will earn 5 points. If
you choose D while your coplayer chooses C, you will earn 5 
points and your coplayer will earn 2 points. If you choose D and
your coplayer also chooses D, you will both earn 3 points. Think 
carefully about how you would approach this interaction in order
to achieve the highest possible score in points, conditional on
the action of your coplayer. Please think step by step before 
making a decision. Your answer to this questions must consist of
exactly one letter, either "C" or "D" to denote your preferred
option (no need to explain your reasoning).
\end{Verbatim}

\section*{Appendix B: Additional Figures}

\begin{figure}[h]
  \subcaptionbox{Prisoner's Dilemma}{\includegraphics[width = 2.7in]{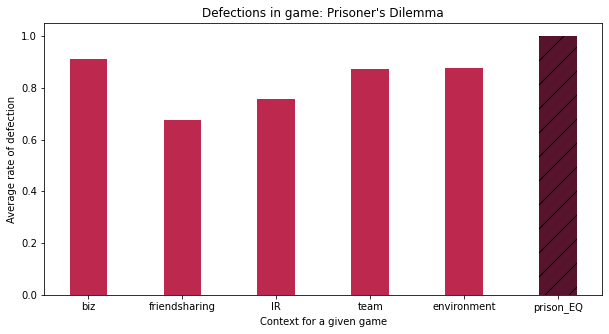}}\quad
  \subcaptionbox{Snowdrift}{\includegraphics[width = 2.7in]{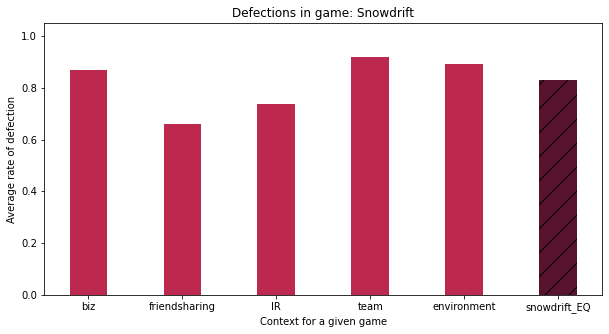}}
  \subcaptionbox{Stag Hunt}{\includegraphics[width = 2.7in]{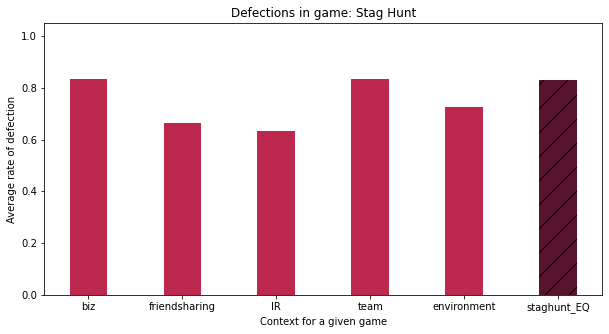}}\quad
  \subcaptionbox{Prisoner's Delight}{\includegraphics[width = 2.7in]{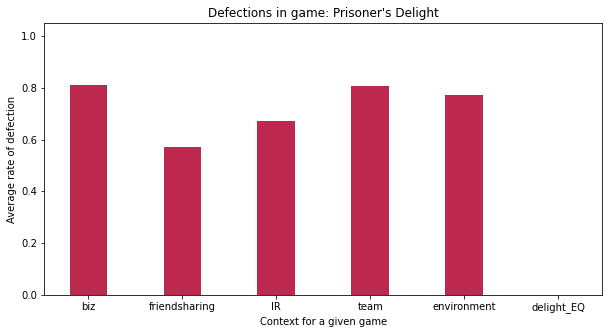}}
  \caption{Bar chart visualization of the propensity to defect or cooperate for each context and for each game using GPT-3.5. In red, the percentage of times the algorithm chose to defect. The dark red striped bar indicates equilibrium values. in the Prisoner's Delight, a rational player would never defect, and thus no bar is displayed. For Stag Hunt and Snowdrift, we indicate as "equilibrium" the probabilities an equilibrium mixed strategy would assign to either action, but both games possess multiple equilibria in pure strategies.}
  \label{fig1}
\end{figure}

\begin{figure}[h]
  \subcaptionbox{Prisoner's Dilemma}{\includegraphics[width = 2.7in]{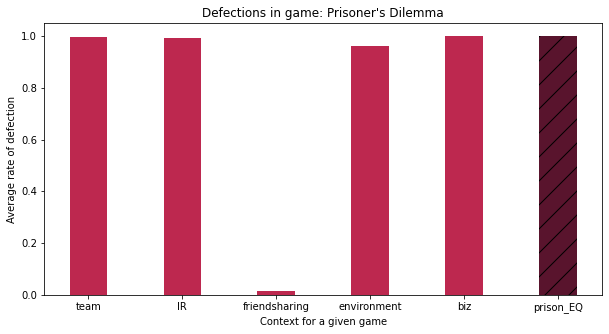}}\quad
  \subcaptionbox{Snowdrift}{\includegraphics[width = 2.7in]{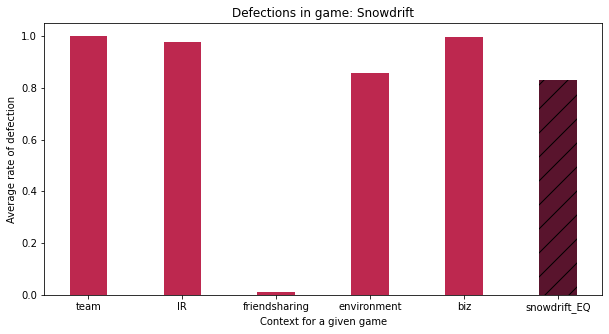}}
  \subcaptionbox{Stag Hunt}{\includegraphics[width = 2.7in]{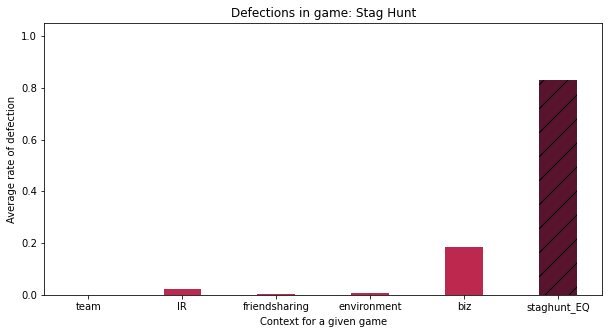}}\quad
  \subcaptionbox{Prisoner's Delight}{\includegraphics[width = 2.7in]{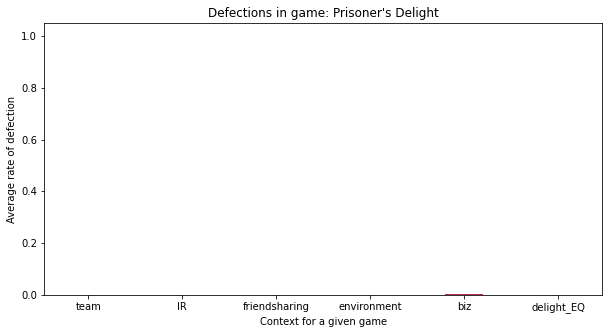}}
  \caption{Stacked bar chart visualization of the propensity to defect for each context and for each game using GPT-4. The methods employed are the same as those described in Figure \ref{fig1}}
  \label{fig4}
\end{figure}

\begin{figure}[h]
  \subcaptionbox{Prisoner's Dilemma}{\includegraphics[width = 2.7in]{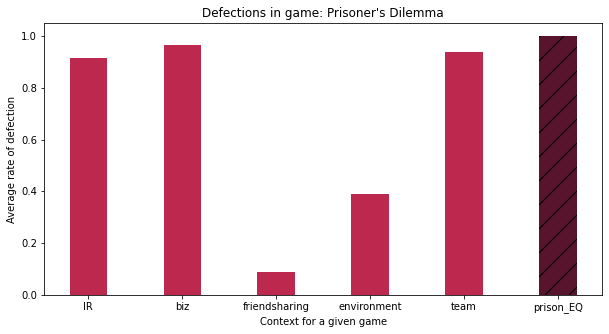}}\quad
  \subcaptionbox{Snowdrift}{\includegraphics[width = 2.7in]{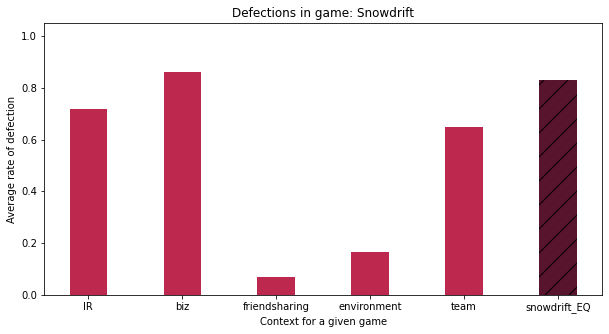}}
  \subcaptionbox{Stag Hunt}{\includegraphics[width = 2.7in]{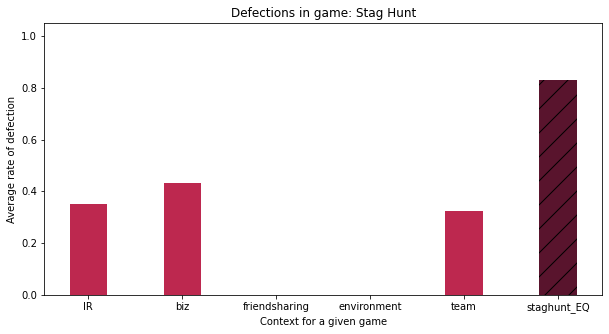}}\quad
  \subcaptionbox{Prisoner's Delight}{\includegraphics[width = 2.7in]{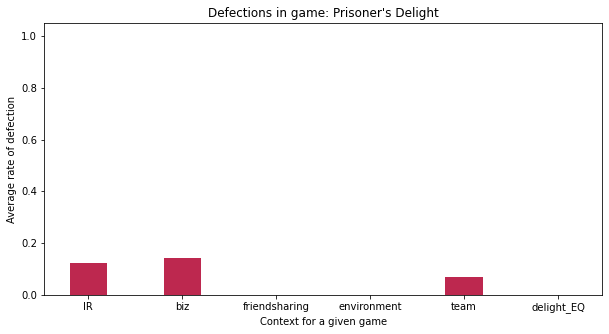}}
  \caption{Bar chart visualization of the propensity to defect for each context and for each game using LLaMa-2. The methods employed are the same as those described in Figure \ref{fig1}}
  \label{figX}
\end{figure}

\end{appendices}

\end{document}